\documentclass[sn-mathphys,Numbered]{sn-jnl}

\usepackage{graphicx}%
\usepackage{amsmath,amssymb,amsfonts}%
\usepackage{amsthm}%
\usepackage{mathrsfs}%
\usepackage[title]{appendix}%
\usepackage{xcolor}%
\usepackage{textcomp}%
\usepackage{manyfoot}%
\usepackage{booktabs}%
\usepackage{algpseudocode}%
\usepackage{listings}%

\usepackage{framed,multirow}

\usepackage{latexsym}

\usepackage{diagbox}
\usepackage{tabularx}
\usepackage{booktabs} 
\usepackage{longtable}

\usepackage{url}
\usepackage{xcolor}
\definecolor{newcolor}{rgb}{.8,.349,.1}

\usepackage{hyperref}

\usepackage[switch,pagewise]{lineno} 

\usepackage{bm}

\usepackage{algorithm}
\usepackage{algorithmicx}
\usepackage{algpseudocode}

\begin{document}

\title[Introducing Anisotropic Fields for Enhanced Diversity in Crowd Simulation]{Introducing Anisotropic Fields for Enhanced Diversity in Crowd Simulation}

\author[1]{\fnm{Yihao} \sur{Li}}
\email{lyh@bit.edu.cn}
\author[1]{\fnm{Junyu} \sur{Liu}}
\author[1]{\fnm{Xiaoyu} \sur{Guan}}
\author[1]{\fnm{Hanming} \sur{Hou}}
\author*[1]{\fnm{Tianyu} \sur{Huang}}
\email{huangtianyu@bit.edu.cn}
\affil[1]{\orgdiv{Key Laboratory of Digital Performance and Simulation Technology}, \orgname{Beijing Institute of Technology}, \orgaddress{\street{Zhong Guan Cun South Street}, \postcode{100081}, \state{Beijing}, \country{China}}}

\abstract{
Large crowds exhibit intricate behaviors and significant emergent properties, yet existing crowd simulation systems often lack behavioral diversity, resulting in homogeneous simulation outcomes. To address this limitation, we propose incorporating anisotropic fields (AFs) as a fundamental structure for depicting the uncertainty in crowd movement. By leveraging AFs, our method can rapidly generate crowd simulations with intricate behavioral patterns that better reflect the inherent complexity of real crowds. The AFs are generated either through intuitive sketching or extracted from real crowd videos, enabling flexible and efficient crowd simulation systems. We demonstrate the effectiveness of our approach through several representative scenarios, showcasing a significant improvement in behavioral diversity compared to classical methods. Our findings indicate that by incorporating AFs, crowd simulation systems can achieve a much higher similarity to real-world crowd systems. Our code is publicly available at https://github.com/tomblack2014/AF\_Generation.  
}


\maketitle

\section{Introduction} 
The utilization of computer algorithms for simulating and modeling human crowds has diverse applications across various fields, including social sciences, virtual reality, visual effects, and game entertainment \cite{11,12,18}. Currently, research focused on simulating crowd behavior, though established for over 20 years, still faces challenges and opportunities for further development, with many studies utilizing artificially designed scenarios as a primary guiding force \cite{20,24,27}. Specifically, most crowd simulation studies employ a two-layered model that integrates global path planning and local navigation \cite{1}. The former identifies broad global routes and imparts behavioral semantics for crowds on a macroscopic level. Meanwhile, the latter entails mapping out intricate local behaviors, supplying immediate and localized motion states for crowds. This approach is efficient in highly deterministic scenarios such as evacuation, formation, and marching \cite{16,18,19}. However, genuine crowds are complex and emergent, and traditional workflows struggle to achieve realistic, general, and effective results owing to the insufficient incorporation of uncertainty modeling within crowd behavior.

\begin{figure*}[tb]
\centering 
\includegraphics[width=\linewidth]{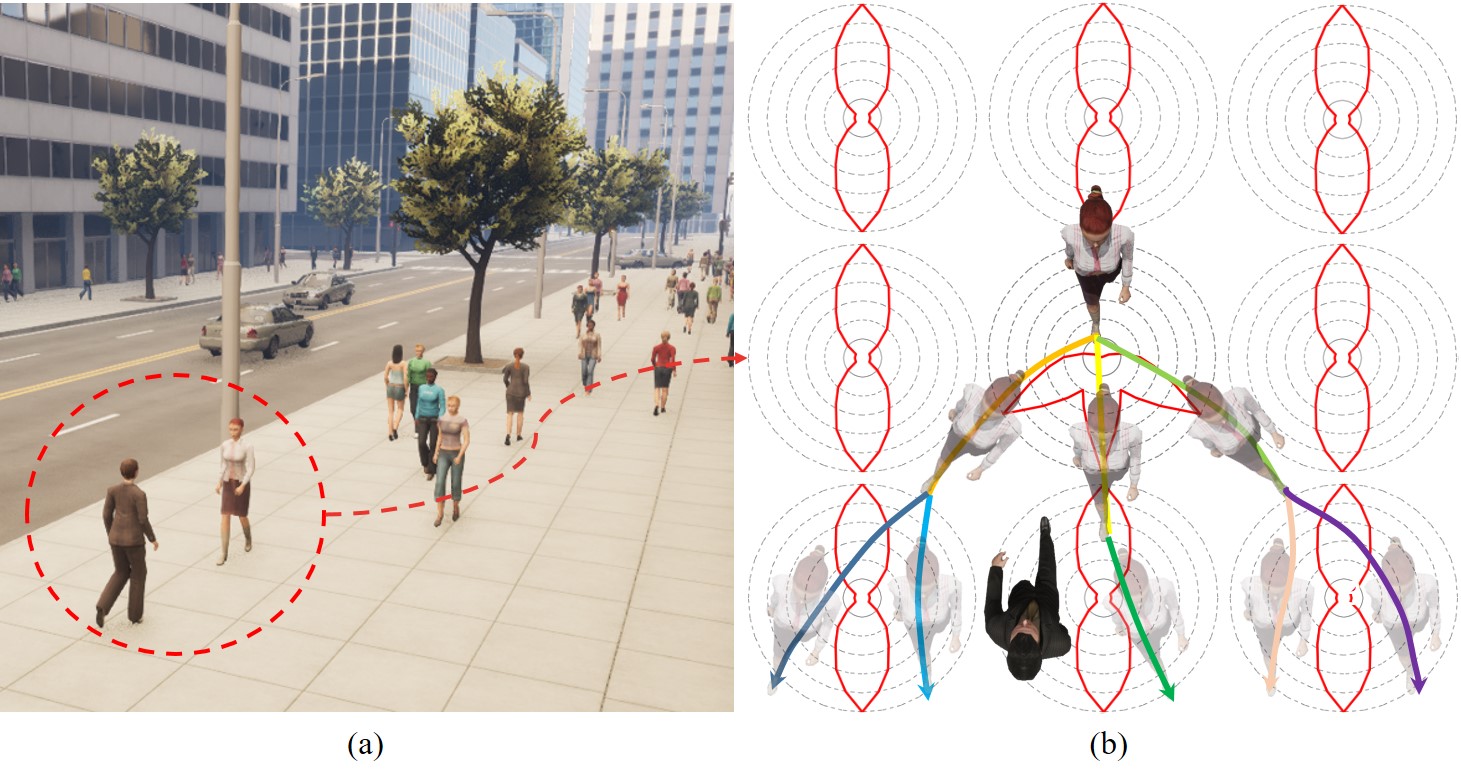}
\hfill \mbox{}
\caption{Potential impact of anisotropic fields on agent motion in a simulated scenario. (a): simulated scene. (b): illustration of an agent's behavioral diversity via AFs via two simulation iterations. Only one behavior occurs in the actual simulation; the result is uncertain.}
\label{fig:teaser}
\end{figure*}

Despite the multitude of related studies conducted over the past two decades, several significant issues persist within the previously mentioned modeling framework. First, while global path planning achieves macro-level planning through global information routing, its implementation relies heavily on users’ definition of crowd behavioral semantics at a macro level. Consequently, the simulated crowd behavior tends to converge significantly, posing a challenge to achieve complex and emergent crowd simulation at the system level. Second, the deterministic nature of local navigation leads to similar behavioral patterns among groups placed in similar environments, making it challenging to capture the complexities observed in real crowds. A considerable portion of previous research employing identical parameters has often led to the replication of nearly identical group motion results. To address these issues, prior research has introduced crowd simulation systems, including crowd motion using patch-based fragmentation \cite{2,3,4}, data-driven crowd trajectory prediction \cite{5,6}, and vector field models that incorporate fluid dynamics \cite{7}. However, a common hurdle encountered is that virtual crowds in simulations are typically limited to executing single and simplistic behavior patterns. Achieving more complex patterns of behavior requires detailed manual grouping, which is both inefficient and yields unnatural results. Consequently, traditional methods can only effectively simulate crowds with simple and deterministic behavior patterns, such as crowds marching or evacuating through crowded corridors \cite{8,9}. Hence, it becomes necessary to introduce an appropriate degree of uncertainty into simulation methods and establish a meaningful connection between macro and micro-level instructions to generate intelligent and diverse results.

Our inspiration stems from the weak correlation observed between individual behavioral patterns and their respective environments. This observation underscores the fact that crowds manifest diverse behavioral patterns in different areas, and even within the same area, different individuals display distinct behavioral patterns. To address the deficiency of diversity semantics in both global path planning and local navigation, we propose the integration of anisotropic fields (AFs) as an anisotropic layer within the crowd system. In contrast to the simple and deterministic macro-level behavioral semantics defined by the navigation field \cite{7}, our model introduces a substantial level of uncertainty into the crowd’s motion behavior within the scene, as illustrated in \autoref{fig:teaser}. The challenge lies in suppressing the dispersion of the simulation results through the innovative introduction of behavioral inertia, thereby enabling the new structure to generate highly complex crowd motion patterns. Furthermore, we have devised a mechanism for generating and self-updating AFs. This mechanism permits the field to be continuously updated using real-time simulation results or data sourced from the real world. Consequently, our work focuses on generating flexible crowd simulation systems that consist of human-like agents capable of exhibiting a wide range of behavioral motion patterns. This approach diverges from the conventional approach of merely directing crowds to perform predefined tasks.

Our primary contributions are summarized as follows:

1.	We introduce a novel anisotropic layer within crowd simulation systems, which is a new structure and does not belong to any classical structure, such as global path planning or local navigation. This addition addresses the absence of behavioral diversity in traditional crowd systems. At its core, our approach utilizes AFs to effectively represent the potential motion tendencies of crowds within the scene using probability distributions.

2.	We design a series of user interfaces that enable real-time modifications to AFs through virtual brush strokes. This user-friendly approach empowers users to create crowd simulation systems that align with their expectations through intuitive sketching.

3.	We present a method for measuring the similarity between crowd systems based on AFs and information entropy. Through experiments, we validate the similarity relationship between the simulated crowd systems generated using our approach and the original real crowd systems.

The remainder of this paper is structured as follows: Section 2 provides an overview of relevant research in various fields related to our study. Section 3 presents the core method based on probabilistic vector fields for crowd simulation and various implementation details. Section 4 describes the experimental quantification methods used in this study. Section 5 details the designed experiments and results. Finally, in Section 6, we summarize the study and discuss potential avenues for future research.

\section{Related Work}

The concept of crowd simulation emerged from early computer research on simulating group behavior patterns \cite{61}. Over the past few decades, crowd simulation methods have become increasingly complex. Reviews and books \cite{11,12,13} provided comprehensive overviews of agent navigation and crowd simulation. Our research focuses on simulating virtual crowds at the macro level using fields with high degrees of freedom and emergent behavior. In the following sections, we discuss the research status in related fields from four perspectives.

\subsection{Field-based Crowd Simulation}

In simulating virtual crowds, researchers often aim to control the movement of large crowds macroscopically to achieve certain behavioral patterns, in addition to explicitly computing the behavior and routes of each agent. Inspired by velocity field systems in fluid mechanics research, researchers have extensively explored simulation methods for crowd flow behavior patterns based on vector fields over the past 20 years. Early works such as flow tiles proposed by Chenney et al. \cite{14} and an interactive algorithm based on navigation fields by Patil et al. \cite{7} filled the gap left by micro-level algorithms like force and velocity-based methods, serving as a supplement at the macro level.

In recent years, field-based crowd simulation methods have shown excellent results in various specific tasks. Toll et al. \cite{15,16} conducted fluid dynamics modeling in high-density crowds, improving their stability while retaining some agent characteristics and achieving easier replication of wave-propagation effects. Montana et al. \cite{17} built on the work by Patil et al.\cite{7} by implementing the simulation and rendering thousands of agents in real-time using the Unreal Engine. Colas et al. \cite{18} introduced interaction fields bound to the agent rather than the scene, to generate a new type of crowd-steering behavior, resulting in enhanced crowd behavior based on agent-based methods in various scenarios.

Real crowds are complex systems that often exhibit high levels of chaos and complexity during operation. Our proposed system based on AFs fully incorporates the ideas of the navigation fields mentioned above. However, the key difference is that our AFs introduce a significant amount of uncertainty into the process to generate emergent results. Simultaneously, our method exhibits similar macroscopic characteristics to the aforementioned methods, resulting in high computational efficiency.

\subsection{Data-driven Crowd Simulation}

Implementing crowd simulation using real crowd data through data-driven approaches is an effective technique. This method simulates virtual crowds by learning the behavioral patterns of real crowds \cite{19,20}. It has the advantage of saving considerable parameter tuning work in crowd systems \cite{21,22,23}. Ren et al. \cite{24} combined physics-based simulation methods with data-driven approaches to achieve a general multi-agent systems simulation method. However, this method is not ideal for simulating rapidly changing crowds. Ju et al. \cite{25} proposed a groundbreaking approach that generates simulated crowds by blending intermediate states of real crowd behavioral patterns. Zhao et al. \cite{26} used clustering to find similar behavioral patterns and employed artificial neural networks to classify agent-perceived input states and generate corresponding outputs. Liu et al. \cite{27} introduced a velocity-based crowd energy metric to measure the similarity between the simulated agent’s velocity and given velocity samples, achieving a new framework for data-driven optimization in crowd simulation. Yin et al. \cite{59} utilized virtual reality (VR) to capture the user’s movement in the virtual space, thereby realizing the construction of crowd behavior data. Through experiments, they verified that realistic group behavior can naturally emerge from the virtual crowd data generated by their method. Data-driven approaches generally rely on the completeness and reliability of data sources\cite{65}, and current datasets struggle to achieve reliable crowd simulation results regarding sample size and scenario diversity.

\subsection{Sketch-based Control of Crowds}

As mentioned in Section 2.1, flow control is based on the definition and manipulation of vector fields. Sketch-based methods provide an intuitive way to achieve this. Vector field generation for crowd simulation mostly relies on sketching. Kang et al. \cite{28} designed detailed stroke-based flow fields and gradient vector flow generation techniques, achieving efficient generation of crowd simulations. Andrea et al. \cite{29} proposed a real-time crowd motion creation method based on immersive sketching in VR scenes, expanding the application of sketching from 2D to 3D space. Sketch-based methods explicitly influence the generation of scenes \cite{30}, global routes \cite{31,32}, navigation grids \cite{33}, and other factors through user-drawn sketches, guiding the generation of crowd simulation results. Additionally, many studies \cite{7,17,18} have generated fields through sketching.

Sketch-based methods for crowd control are generally characterized by high computational efficiency and intuitive interaction. However, owing to the limitations of sketches themselves, these methods produce results that often exhibit fixed and patterned tasks\cite{66}, making it difficult to represent the complexity exhibited by real crowds.

\subsection{Construction of Crowd Simulation Systems and Parameter Tuning}

From the early great breakthroughs of classic crowd simulation systems such as Social Force Model (SFM) and Reciprocal Velocity Obstacle (RVO) \cite{34,35}, the optimization of computer crowd simulation systems has focused on collision avoidance \cite{36,37,38,39} and pathfinding \cite{40,41,42} of agents in local spaces, rather than on the macro-level behavioral semantics of crowds. The effectiveness of these systems frequently relies on extensive parameter tuning, which can have implicit effects on the results. Currently, industrial-level crowd simulation software and crowd formation control systems exist \cite{46,47}. Their usage depends on the expertise of simulation system designers, and the design cycle for a large-scale crowd simulation can take months. To address this issue, some studies \cite{29,48} offer real-time interactive interfaces for parameter editing to achieve instant feedback. Another approach involves automatically adjusting parameters based on text descriptions \cite{49}, controlling randomization \cite{50,51}, and utilizing other methods. Additionally, mathematicians have introduced anisotropy into pedestrian systems to analyze the “optimal” direction of flow-like pedestrians in scenes \cite{52}. Their method provides a computational model for crowd simulation rather than a modeling method itself. However, the results cannot describe the behavioral diversity of complex systems from a crowd perspective. Some latest works\cite{62,63,64} have achieved the heterogeneous construction of agent behaviors through reinforcement learning, generating remarkable effects. However, they did not establish a connection between the agent and the environment, which is the core purpose of the anisotropic layer proposed in this paper.

In summary, existing crowd simulation systems encounter challenges in describing the complex motion and behavioral semantics of more realistic virtual crowds. Achieving a balance between the precise semantic definition of agents, which requires significant manual effort, and the convergence of simplistic behavioral patterns is difficult. Operators often resort to mastering parameter tuning techniques and constructing simulation environments to achieve simulation designs tailored to specific patterned goals.\cite{67} Our study establishes an anisotropic layer within crowd simulation, addressing the absence of behavioral diversity typically lacking in the deterministic path generation of traditional global path planning. This innovative approach effectively represents the macroscopic behavioral semantics of crowd motion using the core concept of AFs and realize the crowd replication in the system level.

\section{Method}

This section presents the representative method for capturing the macroscopic behavioral semantics of crowds, which we refer to as the anisotropic layer. We discuss the fundamental properties and operational mechanisms of the core concept of AFs, as well as various approaches to achieving the automated generation of AFs.

\begin{figure*}[tb]
\centering 
\includegraphics[width=\linewidth]{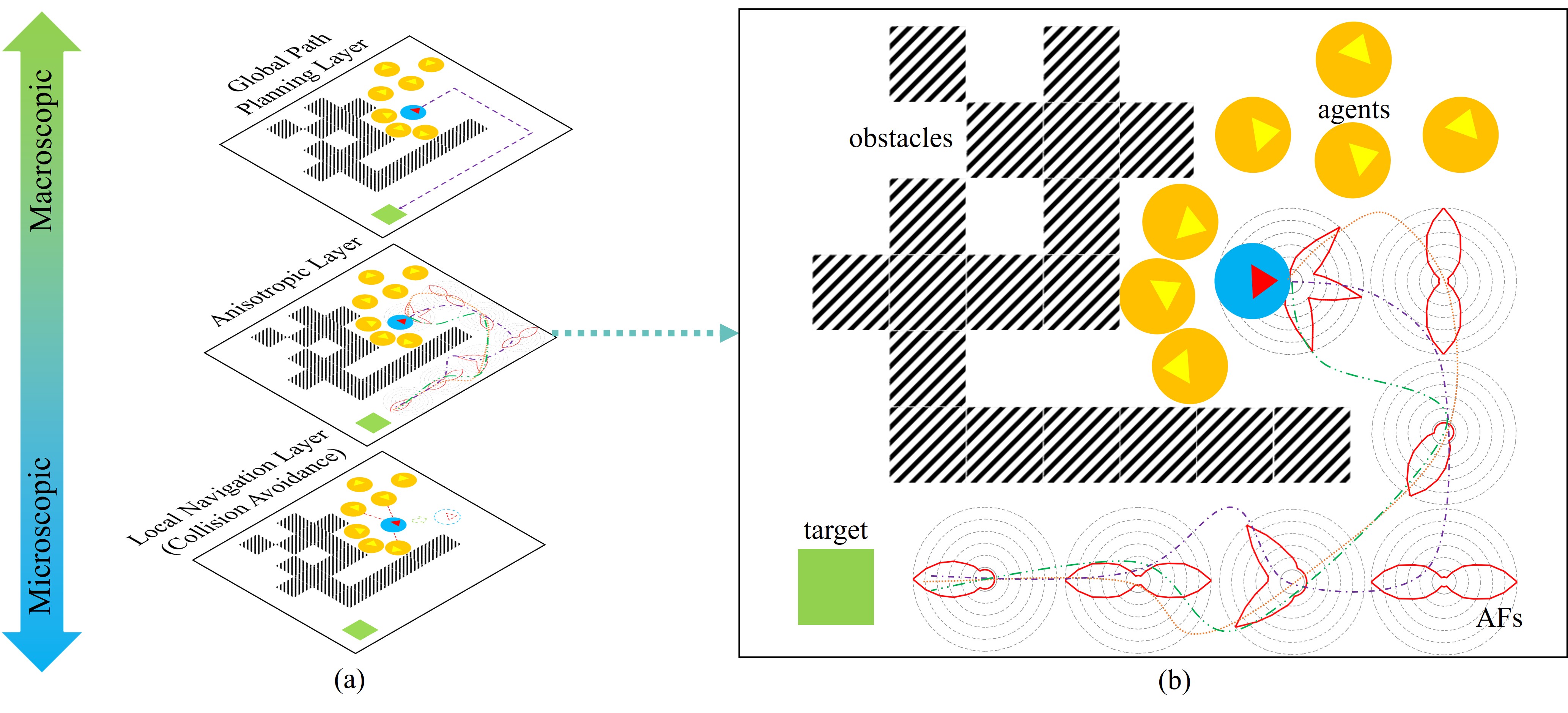}
\caption{Three-layer crowd simulation system after anisotropic layer addition. (a): overview. (b): details of anisotropic layer.}
\label{fig:2}
\end{figure*}

\subsection{Anisotropic Layer}

The anisotropic layer introduced in this study functions differently from both the global path planning and local navigation process in traditional crowd simulation workflows. Within the anisotropic layer, agents display anisotropic behavior steered by a guiding structure in the scene, charaterized by its non-deterministic nature. This process is neither focused on moving agents to specific destinations nor on achieving efficient collision avoidance. The main purpose of the anisotropic layer is to present a deterministic behavioral pattern at the macroscopic level without specifying the precise motion behavior of individual agents. \autoref{fig:2} illustrates the three-layer model of crowd simulation with the inclusion of the anisotropic layer. The anisotropic layer, as proposed, utilizes AFs to enable the execution of its fundamental mechanism.

\subsection{Definition of AFs}

\begin{figure}[tb]
\centering 
\includegraphics[width=0.9\linewidth]{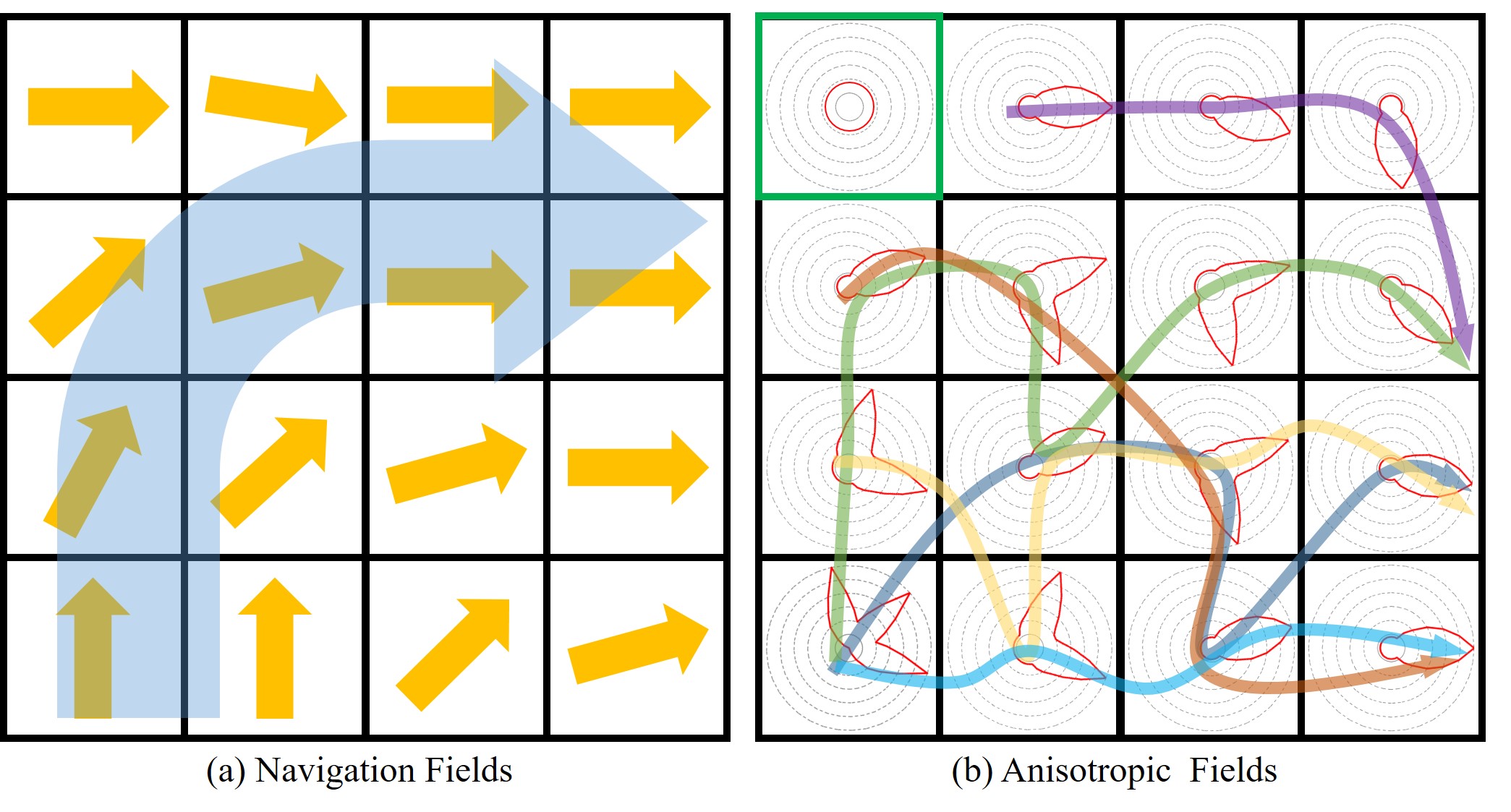}
\caption{Difference between navigation fields and AFs. (a): In traditional navigation fields, agents execute deterministic motion based on the specified navigation directions. (b): In AFs, agents exhibit uncertainty in their movement.}
\label{fig:3}
\end{figure}

In this study, anisotropy, in the context of materials science \cite{53,54}, refers to a property that allows it to change or assume different properties in different directions, as opposed to isotropy. The motion behaviors of agents in a crowd simulation system exhibit properties such as the following: an agent at the same location in a scene may exhibit different movement tendencies, and different agents at the same location in the scene may also exhibit different movement tendencies. This is intuitive. Drawing on previous studies on crowd navigation fields \cite{7} and to address the challenge of motion pattern convergence (resembling isotropic behavior), we propose AFs. An AF is bound to the navigable regions in the scene, covering all reachable area $\Omega$ in the simulation scene. It is discretized to facilitate computation using grid-based division, where $\Omega = G_1 \cup G_2 \cup ... \cup G_n$. $G_i$ and $n$ represent the $i$-th cell and the total number of cells, respectively. In contrast to navigation fields, AFs set a probability distribution $P(G)$ within each cell instead of a single navigation direction. \autoref{fig:3} illustrates the main difference between AFs and navigation fields. By default, $P(G)$ follows a uniform distribution, as shown in the green box in the top-left corner of the AF examples in \autoref{fig:3}. Theoretically, AFs are applicable to navigation in spaces of any dimension. In this study, we discuss a two-dimensional case for simplicity.

\begin{figure}[H]
\centering 
\includegraphics[width=0.9\linewidth]{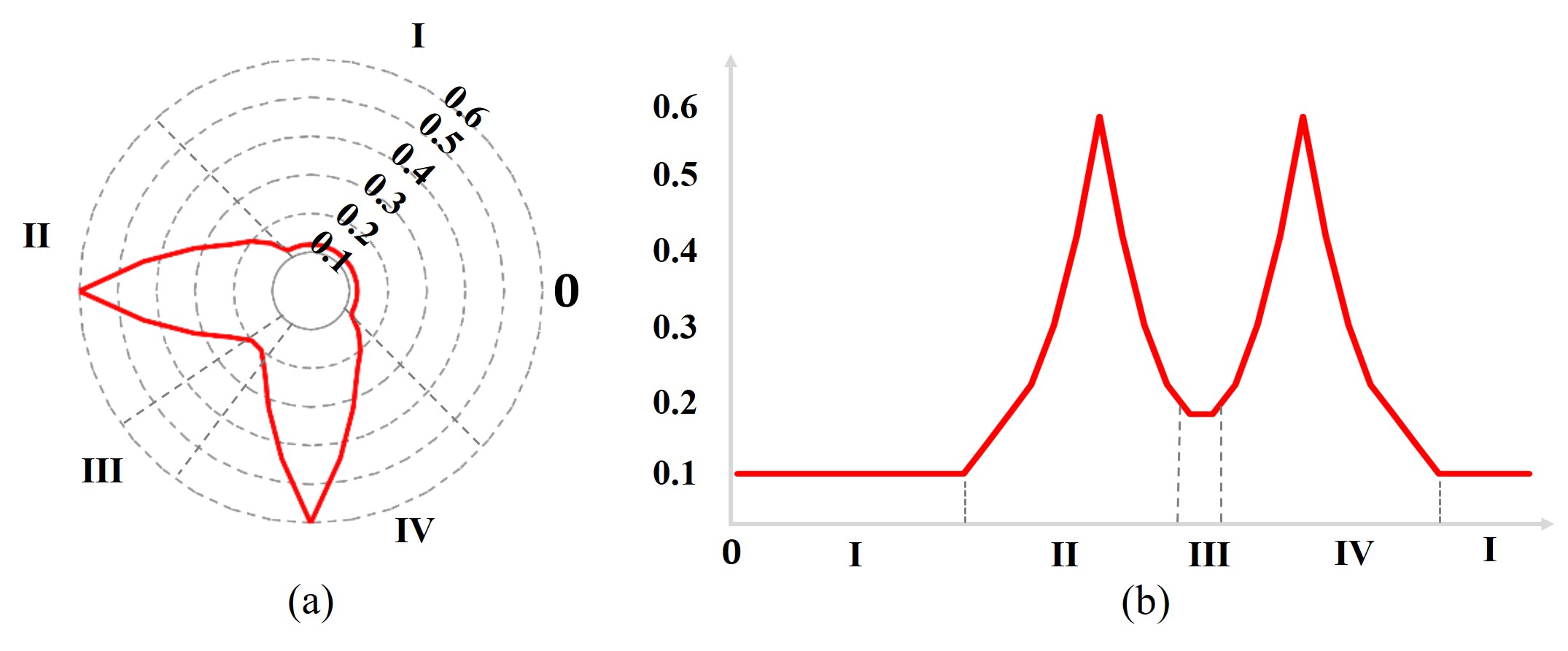}
\caption{ (a) Graphical AF example, (b) Probability function. Agents tend to move toward directions II or IV in this AF region.}
\label{fig:4}
\end{figure}

An AF describes the underlying motion tendencies of agents within the covered grids. Its domain is $[0^\circ,360^\circ)$, which corresponds to the directional tendencies of agents in a two-dimensional scene. \autoref{fig:4}(a) and \autoref{fig:4}(b) illustrate two different representations of the probability distribution for the same position in AF.

The AFs represent the potential motion tendencies of agents in the virtual scenario of the simulation system. Agents are influenced and alter their motion states based on the AFs. In each simulation clock cycle of the system, agents in the scene make direction selections by sampling the distribution corresponding to their $G_i$ in the AF: $P(G_i=\theta)$. The selected result $\theta$ represents the motion tendency. For example, based on the distribution shown in \autoref{fig:4}, an agent is likely to exhibit a motion tendency in either region II or IV, near 180$^\circ$ or 270$^\circ$. In polar coordinates, using $\theta$ as the angle and a constant $\rho$ as the radian, we can compose the acceleration $\bm{a_{AF}}$:

\begin{equation}
\begin{aligned}
&\bm{a_{AF}}=(\theta, \rho)
\end{aligned}
\end{equation}

In the two-dimensional space. The impact of the AFs on agent motion can be combined with classical Euler integration and other models:

\begin{equation}
\begin{aligned}
&\bm{o}\leftarrow\bm{o}+\bm{v}\Delta t\\
&\bm{v}\leftarrow\bm{v}+(k_{AF}\bm{a_{AF}}+\sum_ik_i\bm{a_i})\Delta t
\end{aligned}
\end{equation}

Here, $\bm{o}$ represents the position coordinates of the agent in the environment, $\bm{v}$ is the instantaneous velocity of the agent, $\Delta t$ is the clock cycle of the system simulation, $\bm{a_{AF}}$ is the motion tendency vector generated based on the AF, and $\bm{a_i}$ is the vector representing other influences on the agent, such as tendency or obstacle avoidance. Both $\bm{a_{AF}}$ and $\bm{a_i}$ affect the movement of the agent in the form of acceleration. $k_{AF}$ and $k_i$ are proportionality coefficients. The value of $i$ can be zero, which means agents are only driven by the AFs. We establish the definition of agent speed magnitude by focusing on the agent itself rather than the AF, as in crowd simulation, the pace of movement is often correlated with the agent's role attributes. For instance, the elderly or ill individuals tend to move significantly slower than average young adults.The diversity of agent speed can be determined by a threshold value, denoted as $v_{prop}$, where each agent has an individual $v_{prop}$. When $|\bm{v}|\neq v_{prop}$, the magnitude of $|\bm{v}|$ is adjusted (without changing its direction) based on a coefficient $l$:

\begin{equation}
\begin{aligned}
|\bm{v}| \leftarrow (1-l) \cdot |\bm{v}| + l \cdot v_{prop} 
\end{aligned}
\end{equation}

Based on these rules, the AFs can achieve the property described at the beginning of this section, where agents at the same position in a scene can display diverse motion tendencies, even when comparing different agents at that position. Compared to fixed vector fields in traditional methods, the AFs can generate more complex crowd behavior, making them applicable to a wider range of crowd simulation problems.

\subsection{Behavior Inertia of AFs}

The proposed AF introduces a wide range of possibilities by introducing randomness that affects an object’s movement. However, this can result in frequent fluctuations in crowd behavior within the simulation system because the motion states of agents are updated in each clock cycle to simulate their behavior. For instance, an agent on an east-west road may move eastward in one clock cycle but randomly switch to a westward motion tendency in the next cycle. This erratic behavior leads to unacceptable behavioral jitters.

\begin{figure}[H]
\centering 
\includegraphics[width=0.8\columnwidth]{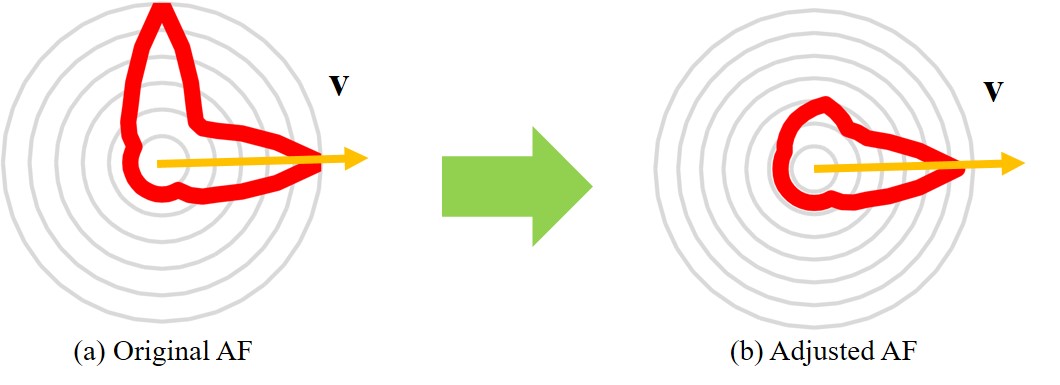}
\caption{Behavioral inertia mechanism.}
\label{fig:5}
\end{figure}

To address this issue, we propose behavior inertia as a solution and promote the convergence of agent behavior. As depicted in \autoref{fig:5}, when an agent calculates its motion tendency using AF, it temporarily modifies the original probability distribution based on its current motion direction $\bm{v}$:

\begin{equation}
p'(\theta)=p(\theta)|cos(\theta-\theta_v)|^\delta
\end{equation}

Here, $p(\theta)$ represents the probability density of AF in the direction $\theta$, and $p'(\theta)$ represents the updated probability density resulting from the adjustment influenced by behavior inertia. $\theta\in[0^\circ,360^\circ)$. $\theta_v$ represents the scalar angle of the agent's current motion direction $\bm{v}$. $\delta$ is a constant used to control the influence of behavior inertia, where a larger $\delta$ represents a stronger influence. Given that $\delta$ weakens $\theta$ which deviates greatly from $\theta_v$, $\delta$ is typically greater than one and theoretically has no upper bound. Through behavior inertia, when agents generate motion tendencies using AF, they tend to actively respond to tendencies that align with their current motion state while disregarding conflicting tendencies, which can effectively solve the possible behavioral fluctuation when the agent switches between grids of adjacent AFs. This characteristic also ensures high coherency in agent behavior within the model, while still allowing diverse and reasonable behavioral variations in scenes that necessitate it, such as intersections. The “temporary” adjustment influenced by behavioral inertia does not directly affect the original AF in the scene but exclusively applies to the agent. Behavior inertia also corresponds to the characteristic that different agents may exhibit different movement tendencies even when occupying the same location.

\subsection{Sketch-based Generation of AFs}

Similar to previous work \cite{17,18,28}, the proposed AFs can be generated using a brush tool for user sketching each curve, denoted as $c$. Each curve has an associated influence weight $\mu_c$ and influence range $\rho_c$. The values of $\mu_c$ and $\rho_c$ are both between 0 and 1. During the sketching session, the user creates a set of curves $C=\{c_1,c_2,...\}$ to indicate the desired probability distribution of AFs at various positions within the target scene. Each curve $c_j$ can be further divided into multiple sub-segments $c_j=\{l_{c_j}^1,l_{c_j}^2,...\}$. These sub-segments are very short, allowing them to be approximated as vector differentials, denoted as $\Delta v$, located at the starting point of each sub-segment. For any $\Delta v$, the influence factor $\tau$ on any cell $G$ within its influence range can be calculated as follows:

\begin{equation}
\tau = \mu_c D_E^{-\varphi}
\end{equation}

Here, $\mu_c$ represents the influence weight of the curve $c$ to which $\Delta v$ belongs, $D_E$ is the Euclidean distance between $\Delta v$ and $p$, and $\varphi$ is a constant that determines the proportion of $D_E$'s influence on $\tau$. To make $\tau$ and $D_E$ negatively correlated, the value of $\varphi$ needs to exceed 1. In this study, we set $\varphi=1.5$. The algorithm for generating AFs based on sketches is Algorithm \autoref{alg:my_algorithm}:

\begin{algorithm}[ht]  
\caption{ Generate AFs at a certain position $G$ based on the sketch lines }  
\label{alg:my_algorithm}  
\begin{algorithmic}[1]  
\Require line set $C$
\Ensure  AF probability distribution $P(G)$ at position $X$
\For{$c_i$ in $C$}
\For{$l_{c_i}^j$ in $c_i$}
\If {length($l_{c_i}^j$, $p$)  $< \rho_{c_i}$}
	\State $\theta$ = angle($l_{c_i}^j$)
	\State $P(X|\theta)$ += $\tau$($l_{c_i}^j$, $X$)
\EndIf
\EndFor
\EndFor
\State Normalize $P(G)$
\end{algorithmic}  
\end{algorithm}

The sketches may sometimes be “self-contradictory”. Therefore, the peak of the motion tendency probability distribution calculated based on the sketch at the same position, can be distributed in opposite directions. This phenomenon is consistent with our daily experience. For instance, on an east-west road, people can move both eastward and westward, rather than only in one direction. Our probabilistic-based model naturally excels at generating more realistic and natural virtual crowds.

\begin{figure*}[ht]
\centering 
\includegraphics[width=\linewidth]{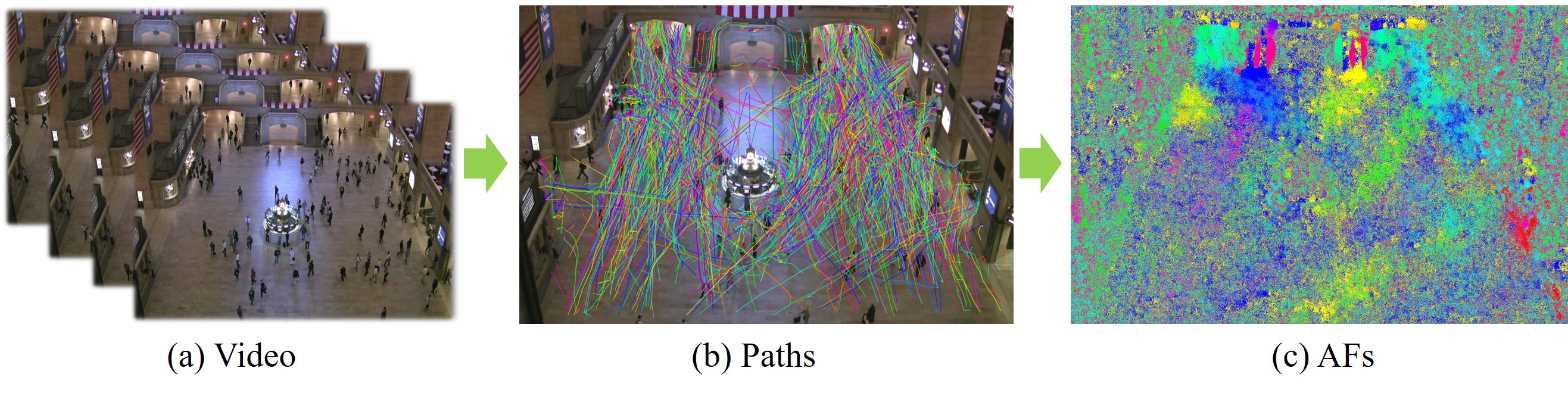}
\caption{Process of AFs generated by crowd video. AFs are represented using the color space-based method mentioned in Section 4.1.}
\label{fig:7}
\end{figure*}

\subsection{Data-driven AF Generation}

By extracting a segment of real-world pedestrian surveillance footage, we employ computer vision techniques (such as optical flow \cite{56}) to derive the motion vectors of pedestrians in each frame of the scene. The optical flow vector $(u,v)$ of the element motion in the video can be calculated as follows:

\begin{equation}
\label{eq5}
I_xu+I_yv+I_t=0
\end{equation}

Here, $I(x,y,t)$ is the target pixel, $I_x=\frac{\partial I}{\partial x}$, $I_y=\frac{\partial I}{\partial y}$, $I_t=\frac{\partial I}{\partial t}$. The optical flow vector of agents and the user’s hand-drawn markings on the AFs are identical and can be explicitly converted between each other. This process can be calculated using the same method as described in Section 3.4. \autoref{fig:7} shows the process of the AF generation method introduced in this section. The representation method of AF differs from the previous one. This color space-based representation method is introduced in Section 4.1. By utilizing the aforementioned method, we generate superimposed AFs or a sequence of AFs that dynamically change with the video. Although we calculate the optical flow and generate AFs in the areas where no one walks in the video(such as walls or pillars), we filter these areas through masks before the actual simulation. In the next section, where we describe our experiments, we demonstrate the significance of utilizing this method to extract generated AFs and generate crowd simulations that closely resemble reality.

\section{Evaluation}

\subsection{Quantification of AFs Based on Information Entropy}

AF can be considered an abstract probabilistic representation of the crowd system, and its distribution characteristics can be effectively characterized using information entropy:

\begin{equation}
\label{eq4}
H=-\sum_{\theta \in \Theta} p(\theta)logp(\theta)
\end{equation}

Here, $\Theta$ represents the complete set of angles in the two-dimensional polar coordinate system, $\theta$ is an arbitrary element, and $p(\theta)$ represents the weighted proportion of $\theta$ in the complete set. Using information entropy, we can calculate the characteristic information of AFs at different locations in the scene. Subsequent visualization and crowd model evaluation are based on the information entropy of AFs.

\begin{figure}[H]
\centering 
\includegraphics[width=0.9\columnwidth]{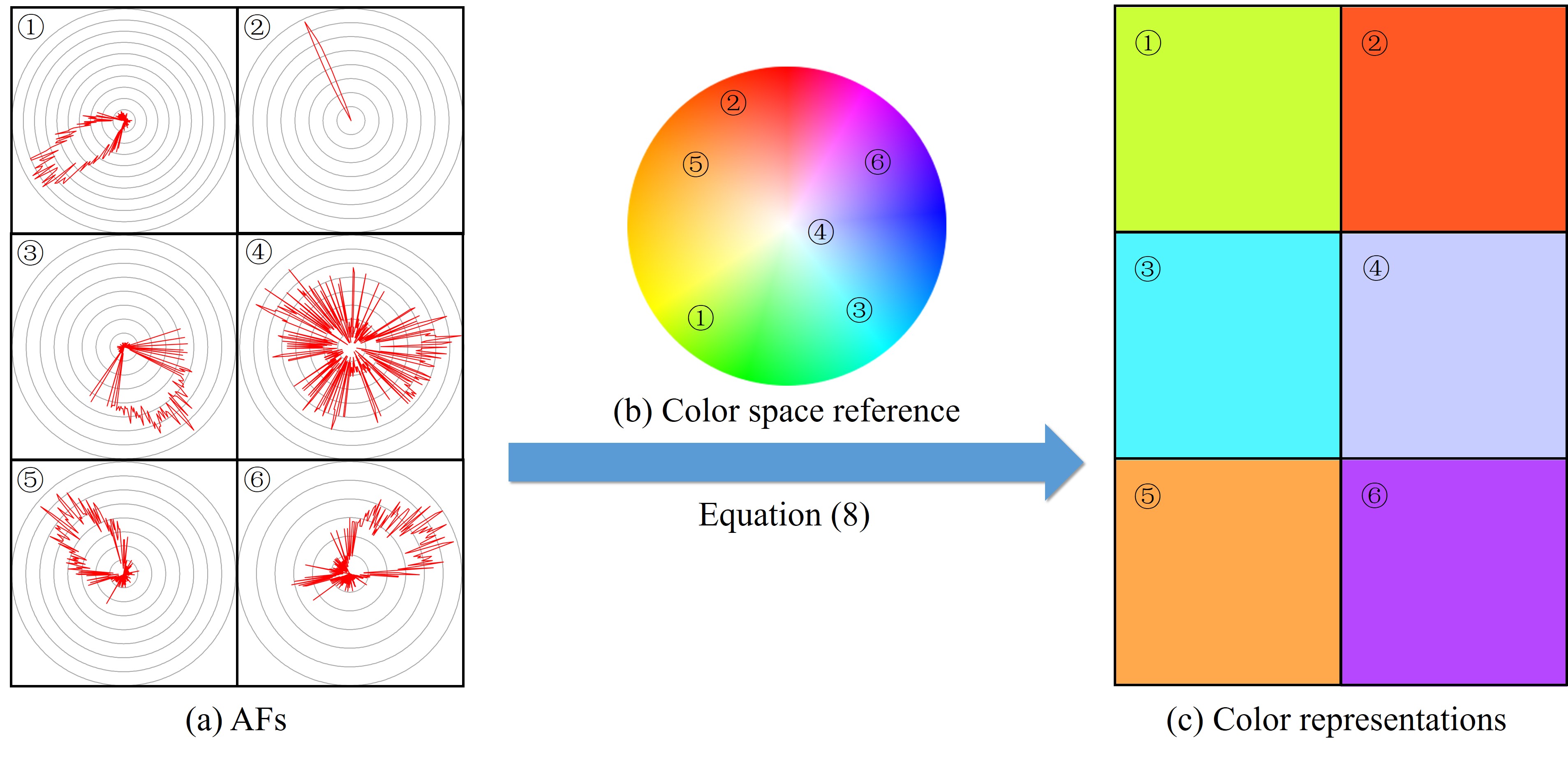}
\caption{Select colors from the color space reference to symbolically represent the AFs}
\label{fig:13}
\end{figure}

In two-dimensional Euclidean space, AFs are distributions on $[0,2\pi)$. To effectively visualize AFs in the subsequent experiments, a color space reference was used to display AFs in different directions, as depicted in the middle of \autoref{fig:13}. For any position in known AFs, the color in the color space reference corresponding to its polar coordinates $(\theta',\rho')$ represents the information of that position:

\begin{equation}
\begin{aligned}
&\theta'=max(P(\theta))\\
&\rho'=1-\frac{H}{H_{min}}\\
&H_{min}=-log\frac{1}{k}
\end{aligned}
\end{equation}

Here, $\theta'$ represents the maximum value of the distribution of the AF in various directions at that position. $H$ is the information entropy at that position, and $H_{min}$ represents the minimum value of theoretical information entropy (under the condition of uniform distribution). Finally, $\rho'$ is inversely correlated with $H$, and $\rho'=0$ when $H=H_{min}$. While this visualization method may lose some information, it intuitively represents the main information in AFs through the color space, assisting readers in gaining a clearer understanding of AFs’ operational mechanism in most cases.

\subsection{Comparison of Macroscopic Similarity in Crowd Systems}

AFs serve as macroscopic abstract expression that not only drives the operation of the crowd system but also represent characteristics of the crowd system evolved from it.

We define the similarity $D$ to characterize the correlation between different AFs as

\begin{equation}
\label{eq6}
D=\frac{1}{2}\sum_{\theta \in \Theta}p(\theta)log\frac{p(\theta)}{p'(\theta)} + \frac{1}{2}\sum_{\theta \in \Theta}p'(\theta)log\frac{p'(\theta)}{p(\theta)}
\end{equation}

Equation \ref{eq6} calculates the Jensen-Shannon divergence\cite{60} of the same position in two crowd systems based on Equation \ref{eq4}, where $p'(\theta)$ represents the probability distribution in the $\theta$ direction at the same position in another crowd system to be compared with $p(\theta)$.

\section{Experiments and Results}

This section presents two different example scenarios, namely multi-intersection channel and data-driven simulation from real crowds, to demonstrate the unique role played by AFs in crowd simulation. For each scenario, the virtual environment was divided into a grid of 10 cm x 10 cm cells, and AFs were distributed in each cell to influence the movement of the crowd. To reflect the differences among various agents, each agent's $v_{prop}$ is assigned a random value set between 1 $\text{m/s}$ and 2 $\text{m/s}$. All visualizations in the experiments were implemented in Unreal Engine 5.0, and the system ran on a computer with the following configuration: Intel(R) Core(TM) i9-10900KF CPU 3.70 GHz, NVIDIA GeForce RTX 3080 GPU, and 64 GB of RAM.

\subsection{Calculation Efficiency}

\begin{figure}[h]
\centering 
\includegraphics[width=0.7\columnwidth]{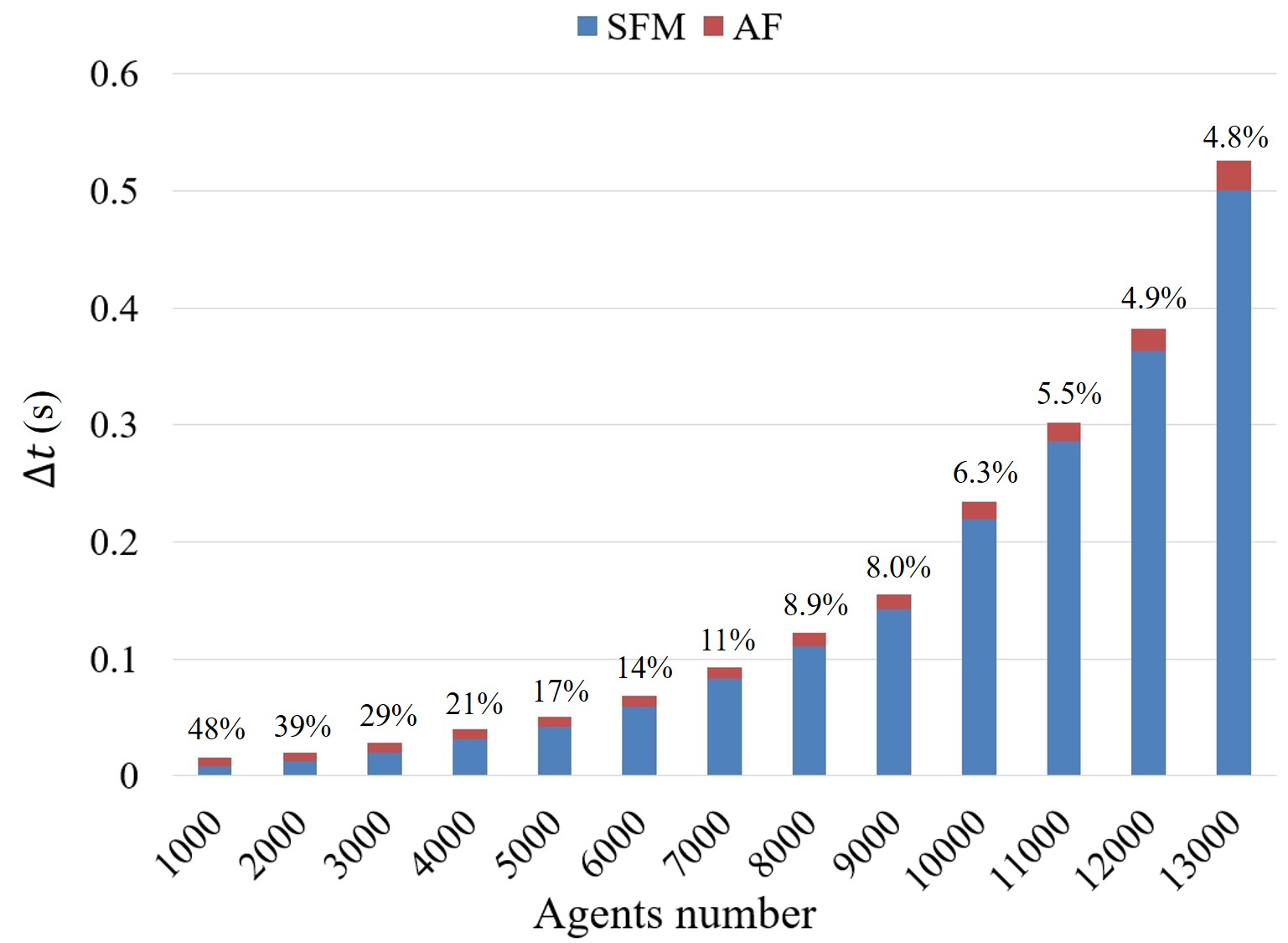}
\caption{Proportion of resource consumption in AF vs. SFM with increasing simulation agents.$\Delta t$ is the time consumed by the algorithm running for one clock cycle.}
\label{fig:8}
\end{figure}

\autoref{fig:8} illustrates the relationship between the computational resource consumption ratio of the proposed method combined with the SFM\cite{34} and the number of simulated agents in crowd simulation. As the count of agents escalates, the proportion of AF in the total computational resource consumption gradually decreases. When the number of agents exceeded 10k, the computational resource consumption outside of AF reached approximately 95\%. In practical applications, if only the macroscopic movement of the crowd is of interest, the proposed method can be used to calculate the crowd movement without running SFM, RVO\cite{34,35}, or other obstacle avoidance methods. With a regular PC, frame rates of over 60FPS can be achieved in crowd simulations at the 10k level.

\subsection{Multi-Intersection Channel Scenario}

\begin{figure*}[h]
\centering 
\includegraphics[width=0.95\linewidth]{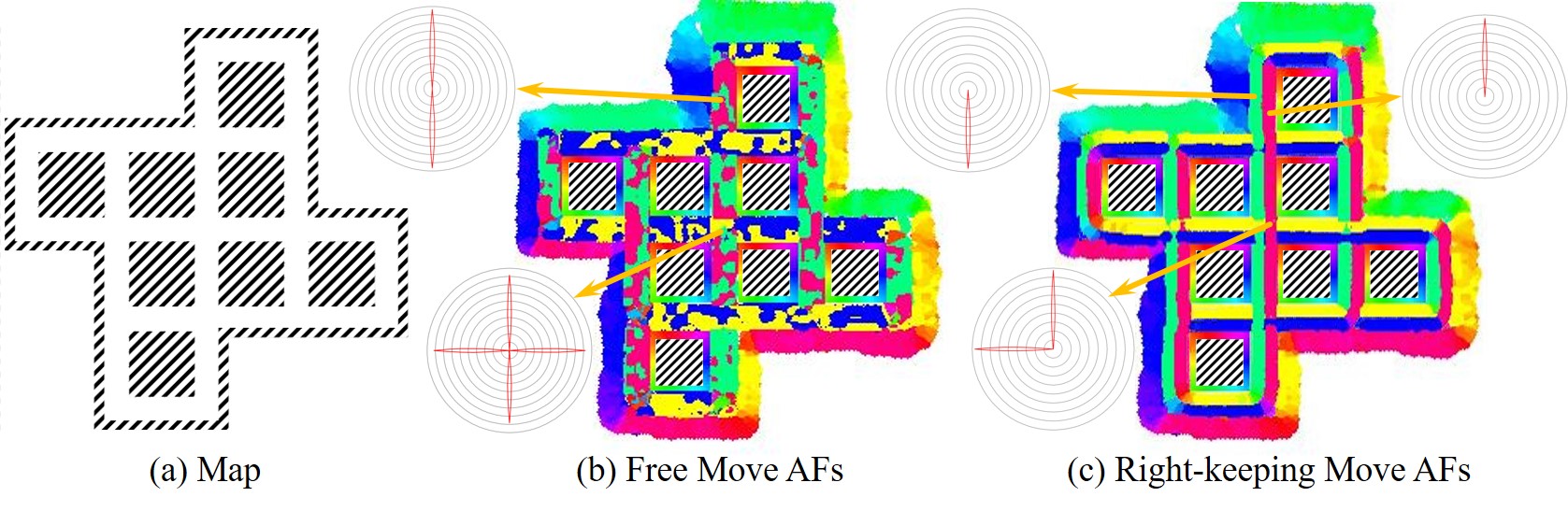}
\caption{Closed-loop channel structure (4 T-shaped intersections, 5 intersections). (a): Structure visualization; (b): Free move AFs distribution (c): Right-keeping move AFs distribution. The radar chart representations of AFs in specific positions are given.}
\label{fig:9}
\end{figure*}

A channel with multiple intersections is a common scenario in crowd simulation. To highlight the uniqueness of our method, we designed a channel scene with multiple crossroads, as shown in \autoref{fig:9}. We set up 400 agents in the environment, with a simulation time of two minutes.

We established two distinct crowd movement “criteria” for this environment, which were created through a combination of automatic and manual annotations, as shown in \autoref{fig:10}. In the free movement scenario, AF was superimposed across all positions within the channel in alignment with the channel’s extension. For example, if the channel extended from east to west, the corresponding positions of the AF exhibited an equal probability distribution in both the east and west directions. At intersections, the AF was uniformly distributed among the different directions based on the number of intersections. Owing to two opposing distributions at the same position and the manual annotations, the AF in the \autoref{fig:9}(b) exhibited a zebra-like striped distribution. It features yellow and blue colors indicating the left-right direction and green and red colors representing the up-down direction (refer to the color space reference in the middle of \autoref{fig:13}). In the free movement scenario (upper portion of \autoref{fig:10}), all agents moved forward or backward within the closed-loop channel, irrespective of their positions within the channel. Consequently, a chaotic crowd movement reminiscent of a bustling marketplace was created. By contrast, the keep-right scenario separated the two directions of AF superimposed along the channel, with all AF distributed on the right side of the channel (consistent with traffic rules in countries such as China, the United States, and Germany). Thus, in the example shown in the lower portion of \autoref{fig:10}, all agents moved towards the right side within the channel, resulting in a more orderly crowd system resembling well-structured urban traffic. The performance comparison of two strategies are shown in \autoref{table:1}. The experimental results verify that large-scale crowd moving on the same side of the road can effectively improve the overall motion efficiency of the whole system. These experimental outcomes were attainable by generating AFs based on real crowd data, and the AF-driven crowd system effectively embodied implicit rules within the crowd system.

\begin{table*}[htbp]
\small
\centering
\caption{Performance comparison of two strategies obtained by simulation}
\begin{tabular}{p{3cm}<{\centering} p{2.3cm}<{\centering} p{2.9cm}<{\centering} p{2.9cm}<{\centering} }
\toprule 
 Strategy&  collision $\downarrow$& velocity variance $\downarrow$& intersection pass $\uparrow$ \\
\midrule 
Free move& 218& 0.0177 & 1212 \\
Right-keeping move& 87&0.0080 &1441 \\
\bottomrule 
\end{tabular}
\label{table:1}
\end{table*}

\subsection{Data-Driven AF Generation and Crowd Simulation}

\begin{figure}[H]
\centering
\begin{minipage}[b]{\textwidth}
\centering
\includegraphics[width=1\linewidth]{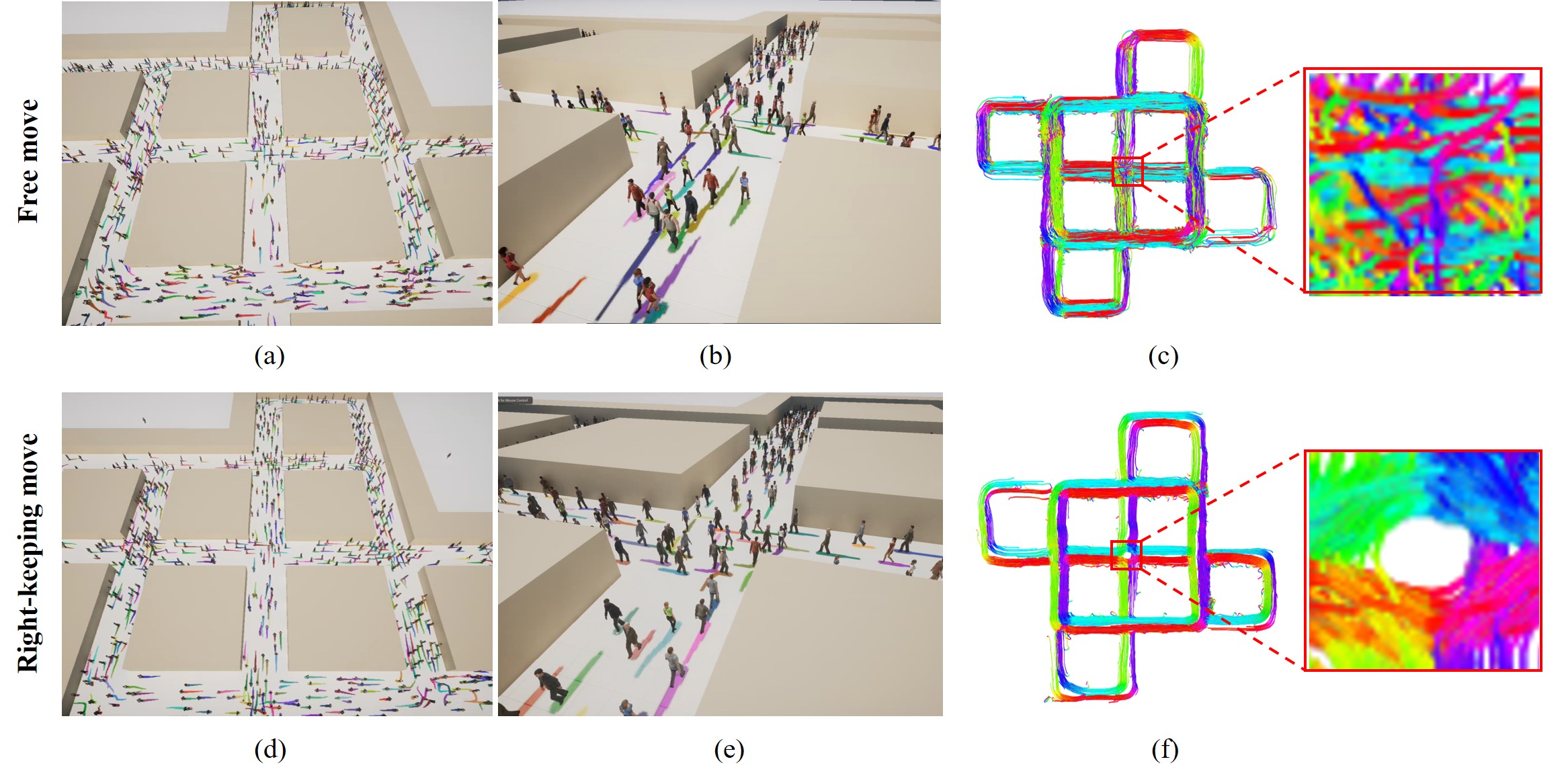}
\caption{Simulation of crowd behavior in a channel environment using AFs with two different strategies. (a)(b)(c): free movement case; (d)(e)(f): right rule case.The AFs corresponding to the two strategies are shown in \autoref{fig:9}.}
\label{fig:10}
\end{minipage}

\vspace{0.3cm}

\begin{minipage}[b]{\textwidth}
\centering
\includegraphics[width=1\linewidth]{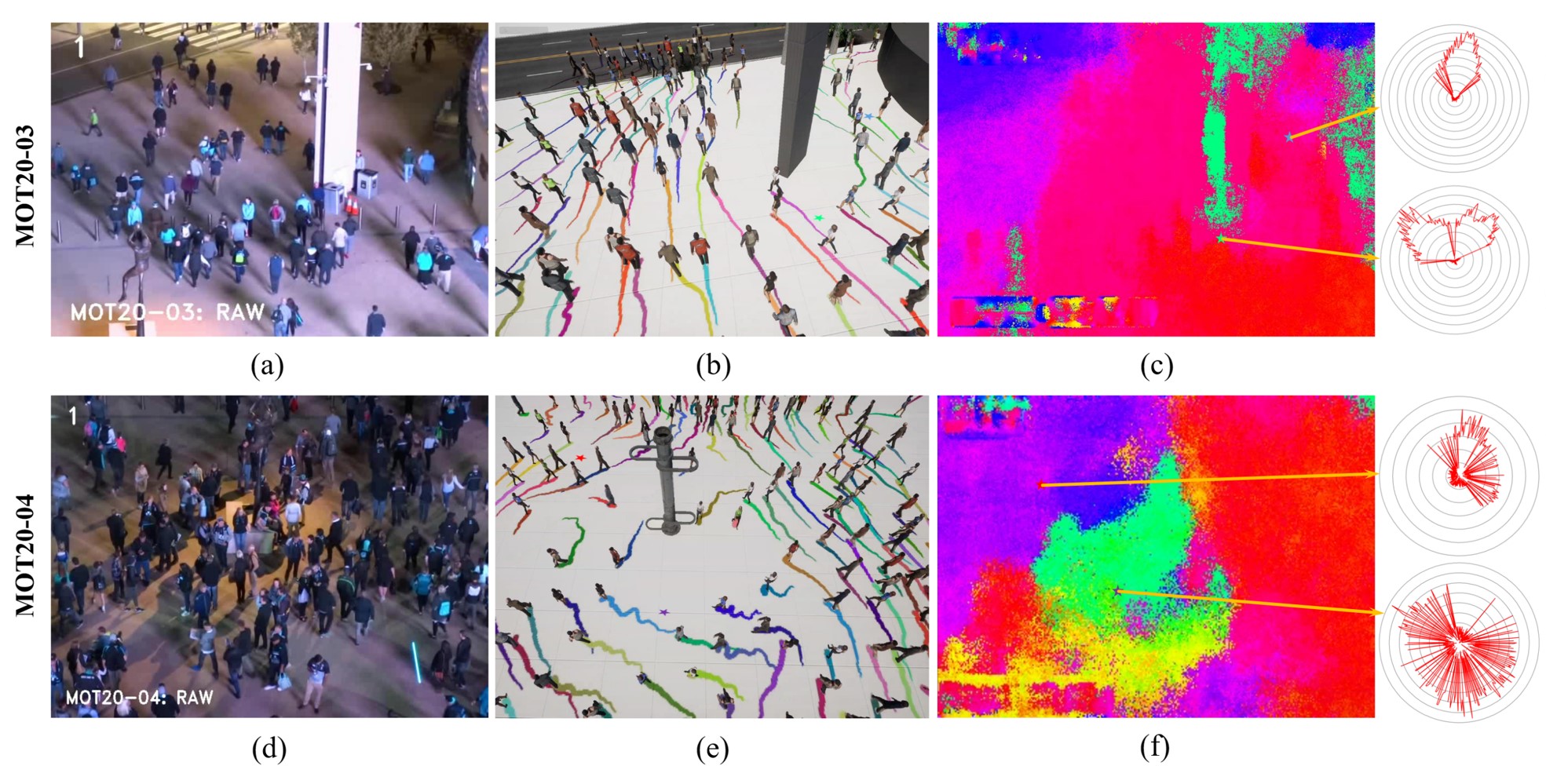}
\caption{AFs generated from MOT20 video and applied to generate crowd systems. (a)(d): Original videos (approximately 90 s, from \cite{57}); (b)(e):AFs-driven virtual crowd systems; (c)(f): AFs generation from extracted accumulated content. The radar chart representations of AFs in specific positions are given. }
\label{fig:11}
\end{minipage}
\end{figure}

In addition to generating AFs through manual annotation, a pivotal application of this study is the creation of AF-driven parallel simulations of virtual crowds that closely resemble real crowds, using crowd-monitoring video data. We leveraged the crowd video dataset from the MOT20 challenge \cite{55} as our data source to achieve a parallel simulation of the crowd within the monitored environment. Initially, we extracted AFs from the crowd-monitoring videos using the method in Section 3.5. The video content is shown in the \autoref{fig:11}(a) and \autoref{fig:11}(d). Based on the extracted AF (\autoref{fig:11}(c) and \autoref{fig:11}(f)), we constructed a proportional simulation environment in Unreal and placed iconic facilities, in the same positions as in the original environment. In this simulation environment, we randomly placed agents with a density similar to that in the original video and drove their movement using AF. The results are shown in \autoref{fig:11}(b) and \autoref{fig:11}(e). 

\begin{figure}[H]
\centering 
\includegraphics[width=0.7\columnwidth]{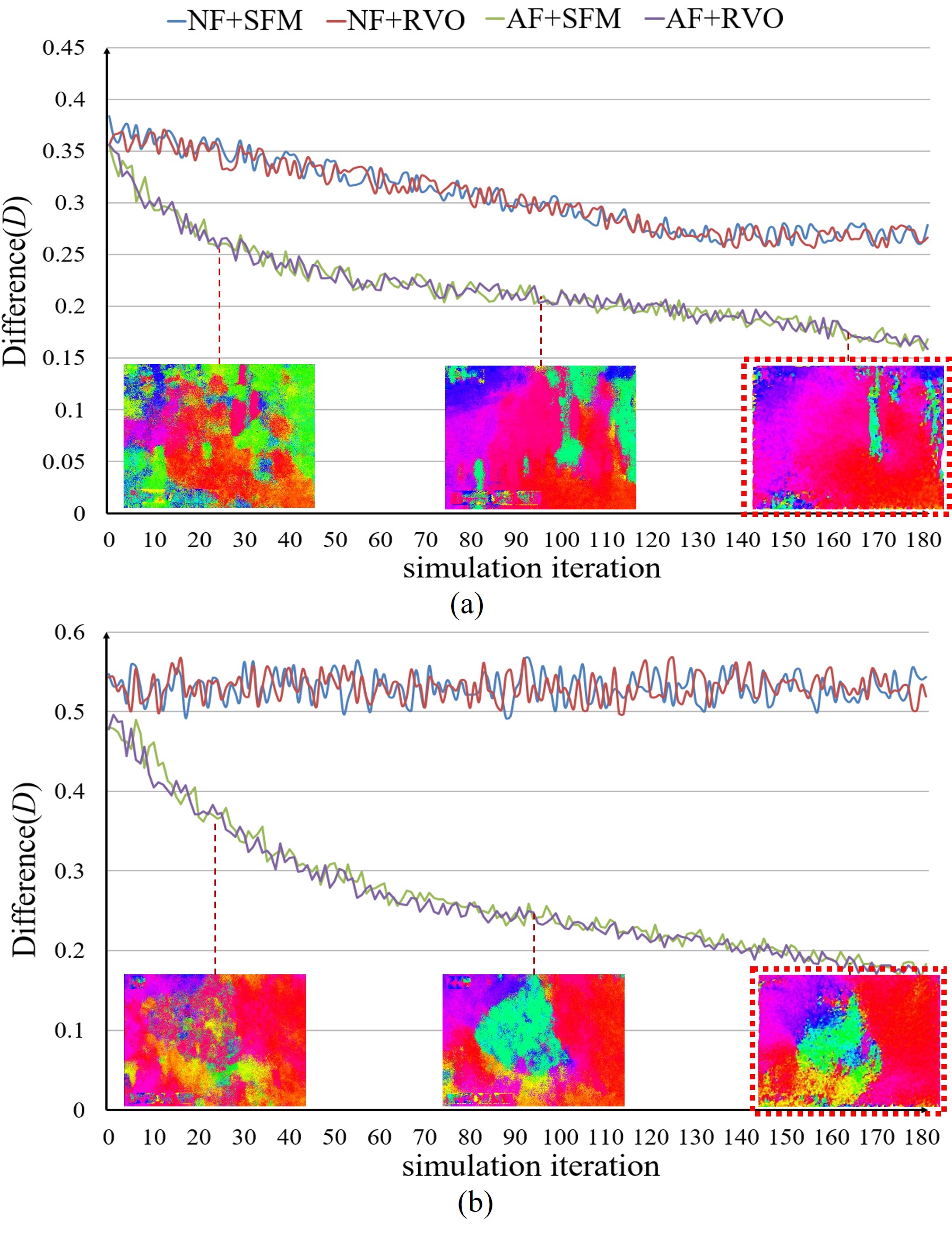}
\caption{Comparison between proposed method and navigation field based crowd system as the simulation time increases based on the $D$ value (Equation \ref{eq6}) of the original crowd video from (a)MOT20-03 and (b)MOT20-04 data.}
\label{fig:12}
\end{figure}

We used the method proposed in Section 4.2 to measure the relationship between the source crowd system and the generated crowd system in \autoref{fig:11}. \autoref{fig:12} presents a comparison of the results implemented by the four combinations of NF+SFM, NF+RVO, AF+SFM and AF+RVO. The two groups of results correspond to MOT20-03 and MOT20-04 respectively. NF refers to the most classic navigation field\cite{7} method and we designed and used it as specified in the original paper. It was observed that as the simulation time increased, the crowd system implemented using our method gradually approached the crowd system in the original video. After 180 iterations, the AF used to represent the two systems became similar, with the distance $D$ decreasing below 0.2(reduced by over 50\% compared to the initial value).It can be observed that the iteratively optimized AF representation of the simulated crowd (the red dashed box in \autoref{fig:12}) demonstrate a remarkable similarity to the AF extracted from the source video (\autoref{fig:11}(c) and (f)). However, the crowd system based on NF exhibited relatively invariant divergence in information entropy as the simulation progressed, resulting in a relatively larger difference in similarity ($D$) compared with the crowd system in the original video. Performance comparison are shown in \autoref{table:2}. It can be observed that our method, while achieving better performance, exhibits comparable runtime speed and shorter build time compared to classical approaches. The crowd simulation method based on field always relies on the manual work of designers. The automated generation method based on video data proposed in this paper can achieve significant improvement in efficiency. Furthermore, another recent related work\cite{18} is challenging to achieve heterogeneous control over hundreds of agents in this context due to its reliance on agent-based field modeling and the lack of automation tools. Therefore, to the best of our knowledge, none of the current methods can achieve the same effect at the macroscopic system level as demonstrated in this experiment.

\begin{table}[htbp]
\small
\centering
\caption{Performance comparison between AF and NF on building crowd systems}
\begin{tabular}{p{1.7cm}<{\centering} p{2.7cm}<{\centering} p{1.7cm}<{\centering} p{1.7cm}<{\centering} p{2.7cm}<{\centering}}
\toprule 
Field & Scenario & $D$ $\downarrow$ & FPS $\uparrow$ & Build Time (s) $\downarrow$ \\
\midrule 
\multirow{2}{*}{NF\cite{7}}&MOT20-03&0.269 & 60+ & 1500 (manual) \\
& MOT20-04&0.531 &60+ & 1800 (manual) \\
\midrule 
\multirow{2}{*}{AF}& MOT20-03&0.165 &58 & 20 (automatic)\\
&MOT20-04 &0.192&56 & 20 (automatic) \\
\bottomrule 
\end{tabular}
\label{table:2}
\end{table}

\subsection{Comparison to similar works}

We have employed seven distinct features to evaluate AFs and compare it with the latest similar work\cite{18} and the most classic work\cite{7}, as summarized in \autoref{table:3}. While all three methodologies exhibit advantages in terms of computational efficiency and integration with steering techniques, the method introduced in this paper stands out with its unique capabilities in utilizing data (including videos and trajectories) for achieving system-level replication. This distinctive aspect underscores the potential of our approach in addressing complex system replication tasks.

\begin{table}[htbp]
\small
\centering
\caption{Judged by some fearures}
\begin{tabular}{p{3.6cm}<{\centering} p{2.6cm}<{\centering} p{2.6cm}<{\centering} p{2.5cm}<{\centering}}
\toprule 
Feature&NF\cite{7}&IF\cite{18}&AF \\
\midrule 
Sketch-based&\checkmark&\checkmark &\checkmark \\
Integratable& \checkmark&\checkmark &\checkmark \\
Efficient& \checkmark&\checkmark &\checkmark \\
Stackable& &\checkmark&\checkmark\\
Data-driven& & &\checkmark\\
System Replication& & &\checkmark\\
Heterogeneous& & &\checkmark\\
\bottomrule 
\end{tabular}
\label{table:3}
\end{table}

\section{Conclusion and Future Work}

This study introduced the use of AFs as a novel approach to describe the randomness and emergent behavior within crowds. This approach serves as a complementary method to traditional agent-based methods for simulating crowd systems, and it maintains backward compatibility with the navigation fields from which it was inspired. AFs are generated through intuitive sketches, empowering users to design crowd systems with specific characteristics. Alternatively, AFs can be derived from crowd trajectory data extracted from videos, resulting in the creation of digital crowd twins that closely resemble real-world behavior. Unlike conventional methods, the simulated crowd systems in this study do not rely on precise mathematical modeling of individual entities. Instead, they mimic real crowds at a macro level, where even subtle or absent changes in initial conditions can lead to entirely distinct simulation outcomes—a phenomenon akin to the butterfly effect observed in chaotic systems \cite{58}. In scenarios with thousands or more agents, the proposed method has special advantages in generating groups of agents with certain behavioral patterns, such as replicating the crowd dynamics in a park.

However, there are limitations to the approach presented here. Although AFs complement navigation fields and significantly broaden the scope of simulation processes and outcomes, they remain approximations rather than exact representations of objective reality. Additionally, the simulation system proposed in this study does not account for individual differences or the cognitive agency of agents. Instead, it focuses on comparing and approximating crowd systems at a macroscopic level. In future work, we will focus on advancing the application of AFs within machine learning generative models, such as generative adversarial networks and diffusion models, by harnessing their capability to generate crowd systems based on real crowds.

\bibliography{template}

\end{document}